&pdflatex

\documentclass[12pt,prd,aps,amssymb,amsmath,tightenlines,showpacs]{revtex4}
\usepackage{graphicx}

\begin{document}

\title{Complex Trajectories in a Classical Periodic Potential}
\author{Alexander G. Anderson$^a$}
\email{aganders@wustl.edu}
\author{Carl M. Bender$^b$}
\email{cmb@wustl.edu}
\affiliation{$^a$Physics Department, Washington University, St. Louis, MO, USA\\
$^b$Department of Physics, Kings College London, Strand, London WC2R 1LS,
UK\footnote{Permanent address: Department of Physics, Washington
University, St. Louis, MO 63130, USA.}}

\begin{abstract}
This paper examines the complex trajectories of a classical particle in the
potential $V(x)=-\cos(x)$. Almost all the trajectories describe a particle that
hops from one well to another in an erratic fashion. However, it is shown
analytically that there are two special classes of trajectories $x(t)$
determined only by the energy of the particle and not by the initial position of
the particle. The first class consists of periodic trajectories; that is,
trajectories that return to their initial position $x(0)$ after some real time
$T$. The second class consists of trajectories for which there exists a real
time $T$ such that $x(t+T)=x(t)\pm2\pi$. These two classes of classical
trajectories are analogous to valence and conduction bands in quantum mechanics,
where the quantum particle either remains localized or else tunnels resonantly
(conducts) through a crystal lattice. These two special types of trajectories
are associated with sets of energies of measure 0. For other energies, it is
shown that for long times the average velocity of the particle becomes a
fractal-like function of energy.
\end{abstract}

\date{\today}

\maketitle

\section{Introduction}
\label{s1}
Recently, there has been a substantial research effort whose aim is to extend
classical mechanics into the complex domain \cite{r1,r2,r3}. Intriguing
analogies with quantum mechanics emerge when conventional classical mechanics is
generalized to complex classical mechanics. For instance, complex trajectories
of particles under the influence of double-well potentials display
tunneling-like behavior when the classical particle has complex energy
\cite{r4,r5}.

In previous work, complex trajectories in periodic potentials were studied
numerically \cite{r6}. These numerical investigations identified two types of
motion for a complex classical particle in a periodic potential and found a
striking analogy with the behavior of a quantum particle in a periodic
potential. Under the influence of the potential $V(x)=-\cos(x)$, a classical
particle having complex energy appears either to remain localized or to move in
one direction from site to site \cite{r6}. These types of motion are analogous
to the valence-band and the conduction-band behavior of a quantum particle in a
periodic potential.

A recent paper showed that the complex classical equations of motion for a
quartic potential can be solved analytically in terms of doubly periodic
elliptic functions \cite{r7}. This paper uses a similar procedure to study
analytically the classical motion of a particle in the potential $V(x)=-\cos(x
)$. The analytic solution allows us reexamine the cosine potential, which until
now was only examined numerically. In agreement with earlier research, our
analysis in this paper leads us to conclude that there exist two special sets of
energies that give rise to these two kinds of classical behavior, and for these
energies we find that this classical behavior does not depend on the initial
position $x(0)$ of the particle. For energies in the first set the classical
trajectories $x(t)$ are periodic; that is, they satisfy $x(t+T)=x(t)$ for some
real $T$. Energies from the second set give quasiperiodic trajectories with the
property that there is a real time $T$ such that $x(t+T)=x(t)\pm2\pi$.

However, this work finds a previous claim about the periodic potential to be
false. Earlier numerical work suggested that when the trajectory is classified
as a function of the energy of the particle, one observes complex energy bands
of nonzero thickness, with some bands giving rise to localized motion and others
resulting in conducting motion \cite{r6}. This paper shows that this observation
of continuous bands of nonzero thickness was an artifact of tracking the motion
of the particle for a limited amount of time. While accumulating numerical error
tends to limit the time that a trajectory can be followed numerically, our new
analytic solution enables us to predict the behavior of a particle for times
that are arbitrarily large. We find that as we increase the time $t$ that the
particle is observed, the behavior of the particle as a function of energy
becomes more complicated. As $t\to\infty$, the average velocity as a function of
the energy of the particle converges to a fractal-like function that is nonzero
for a set of measure zero of energies and zero for all other energies.

The paper is arranged as follows: In Sec.~\ref{s2} we derive an analytic
solution to the complex classical equations of motion for the periodic potential
$V(x)=-\cos(x)$. Next, in Sec.~\ref{s3} we examine the two types of special
complex trajectories and describe the sets of complex energies associated with
each kind of trajectory. Then, in Sec.~\ref{s4} we show how a simple function of
the energy predicts the hopping of the classical particle from well to well. In
Sec.~\ref{s5} we use our new analytic work to evaluate previous research about
the band-like structure of the complex-classical periodic potential. We also
examine the average velocity of the particle as a function of the energy as the
observation period tends to infinity. We give some concluding remarks in
Sec.~\ref{s6}, and in the Appendix we explain our methods in greater detail. 

\section{Analytical Solution for the Periodic Potential}
\label{s2}

The solution to the equation of motion for complex particle trajectories
involves several standard elliptic functions \cite{r8,r9,r10}. The {\it elliptic
integral of the first kind} is given by
\begin{equation}
z=F(\phi,m)=\int_0^\phi\frac{dt}{\sqrt{1-m\sin^2 t}}.
\label{e1}
\end{equation}
Then, the {\it Jacobi amplitude} $\phi={\rm am}(z|m)$ is defined as the inverse
of the elliptic integral of the first kind:
\begin{equation}
{\rm am}(z|m)=\phi\equiv F^{-1}(z,m).
\label{e2}
\end{equation} 
The Jacobi amplitude is a pseudo-periodic function with respect to its first
argument,
\begin{equation}
{\rm am}[z+2r K(m)+2is K(1-m)|m]={\rm am}(z|m)+r\pi
\label{e3}
\end{equation}
for integers $r$ and $s$; $K(m)\equiv F(\pi/2,m)$ is the {\it complete elliptic
integral of the first kind}.

In terms of these functions, we can solve exactly Hamilton's equations of motion
for a particle in the periodic potential $V(x)=-\cos(x)$ in which the position
and momentum of the particle is complex. Hamilton's equations have one integral
of the motion that expresses the conservation of energy. We scale out
unnecessary constants to obtain
\begin{equation}
\frac{1}{2}\left(\frac{dx}{dt}\right)^2-\cos(x)=E.
\label{e4}
\end{equation} 
Using a double-angle trigonometric identity and separating variables, we get 
\begin{equation}
\sqrt{2}\,t=\int_{x(0)}^{x(t)}\frac{dx}{\sqrt{E+1-2\sin^2(x/2)}}.
\label{e5}
\end{equation}
We then substitute $m=2/(E+1)$ and $u=x/2$:
\begin{equation}
t\sqrt{2E+2}+C=\int^{u(t)}\frac{2du}{\sqrt{1-m\sin^2 u}}.
\label{e6}
\end{equation}
From the definition of the Jacobi amplitude, we can then write
\begin{equation}
x(t)=2{\rm am}(at+C|m),
\label{e7}
\end{equation} 
where $a=\sqrt{(E+1)/2}$, $m=2/(E+1)$, and the complex integration constant $C$
is determined by the initial position of the particle. This is the general
solution because it can incorporate any initial position and velocity.

We emphasize that the constants $a$ and $m$ are just functions of the energy,
while $C$ depends on the initial position. Thus, the analytic solution makes it
plausible that the type of motion is dependent primarily on the energy and not
on the initial position of the particle.

Using the pseudo-periodicity of ${\rm am}(z|m)$, we can identify special
energies that give rise to periodic or quasiperiodic trajectories. Suppose that
the path of a particle is periodic (modulo $2\pi$) with period $T$. Then for
some integers $r$ and $s$, we get
\begin{equation}
aT=2K(m)r+2iK(1-m)s.
\label{e8}
\end{equation}
Because $T$ is a real number, we divide by $a$ and take the imaginary part to
get 
\begin{equation}
\frac{s}{r}=\frac{{\rm Im}[K(m)/a]}{{\rm Im}[iK(1-m)/a]}\equiv R(E),
\label{e9}
\end{equation}
where $m=2/(1+E)$ and $a=\sqrt{2+2E}$. Observe that the right side $R(E)$ of
(\ref{e9}) is a function of the energy $E$.

The fact that the ratio $R(E)=s/r$ depends only on $E$ is fundamental to our
discussion of the complex trajectories of classical particles. We will show that
the trajectories of particles with energies $E$ such that $R(E)$ is rational
give rise to a stable time evolution. We refer to the energies that give rise to
this special behavior as {\it classical eigenenergies}. We will show that the
set of eigenenergies has measure zero in the set of all complex numbers. It is
worth noting that if the energy $E_0$ satisfies (\ref{e9}), then we can find $T$
by reducing the fraction in (\ref{e9}) to lowest terms to obtain the numbers $s$
and $r$. We then substitute these numbers into (\ref{e8}) to solve for $T$.
Choosing $E$ so that (\ref{e9}) holds guarantees that $T$ will be a real number.

This analysis reveals a numerical subtlety. If we use (\ref{e3}), we get $x(t+T)
=2\pi r+x(t)$. However, when we solve the equations of motion numerically, we
find that the solutions come in two types, one for which $x(t+T)=x(t)$ (even if
$r\neq0$) and another for which $x(t+T)=2\pi k+x(t)$ with $k=\pm 1$ (even if $r
\ne \pm 1$). Evidently, the appearance of these two different kinds of solutions
is related to taking the functional inverse of the integral in (\ref{e6}).

\section{Description of Classical Trajectories}
\label{s3}

\begin{figure}
\begin{center}
\includegraphics[scale=1.45]{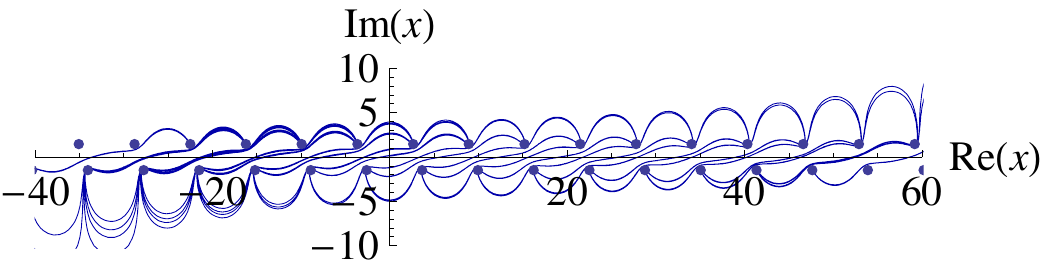}
\end{center}
\caption{A typical trajectory of a particle governed by the cosine potential.
Here, $E=1.99+i$ and $x(0)=0.1 i$. This particle does not exhibit periodic or
quasiperiodic behavior; it just hops at random from site to site.}
\label{F1}
\end{figure}

This paper focuses on the special periodic and quasiperiodic trajectories of the
cosine potential, but it is useful first to examine a typical nonperiodic
trajectory. Figure \ref{F1} displays such a trajectory beginning at $x(0)=0.1i$.
The particle in this figure hops randomly from site to site without exhibiting
any periodic or quasiperiodic behavior.

\begin{figure}
\begin{center}
\includegraphics[scale=1.2]{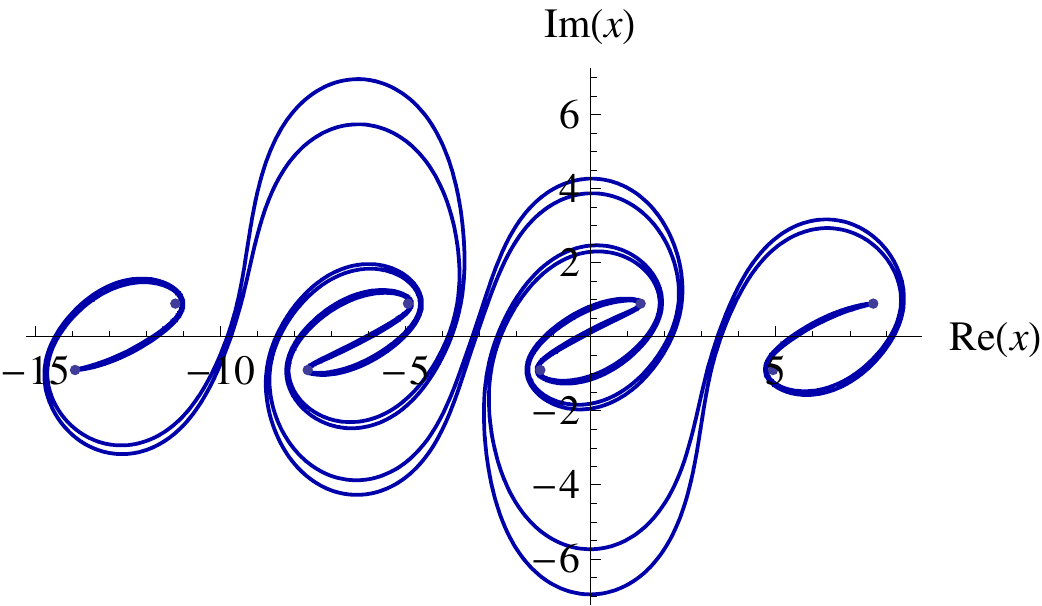}
\end{center}
\caption{Periodic trajectory of a particle of complex energy $E=-0.30905+i$. The
particle starts at $x=0.2i$ and the value of the ratio $R(E)$ is $17/20$.
Because the $R(E)$ is reduced to lowest terms and the integer $17$ in the
numerator is odd, the trajectory $x(t)$ satisfies the phenomenological rule that
$x(t+T)=x(t)$. If the values of the numerator and denominator are increased, the
trajectories get progressively more complicated.}
\label{F2}
\end{figure}

\begin{figure}
\begin{center}
\includegraphics[scale=1.4]{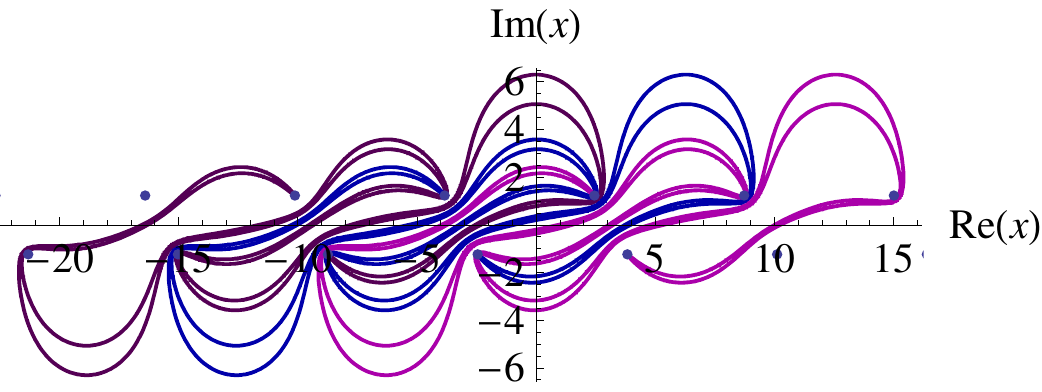}
\end{center}
\caption{Periodic trajectories of a particle of energy $E=1.43696+i$. Again, the
trajectories are periodic because the numerator in the ratio $R(E)=5/12$ is odd.
Three trajectories are shown, one starting at $-2\pi+0.2i$, a second starting at
$0+0.2i$, and a third starting at $2\pi+0.2i$. (These trajectories are purple,
blue, and magenta in the electronic version.) Note that if a trajectory is
translated by $2\pi$ to the left or to the right, it still does not intersect
another trajectory.}
\label{F3}
\end{figure}

\begin{figure}
\begin{center}
\includegraphics[scale=1.25]{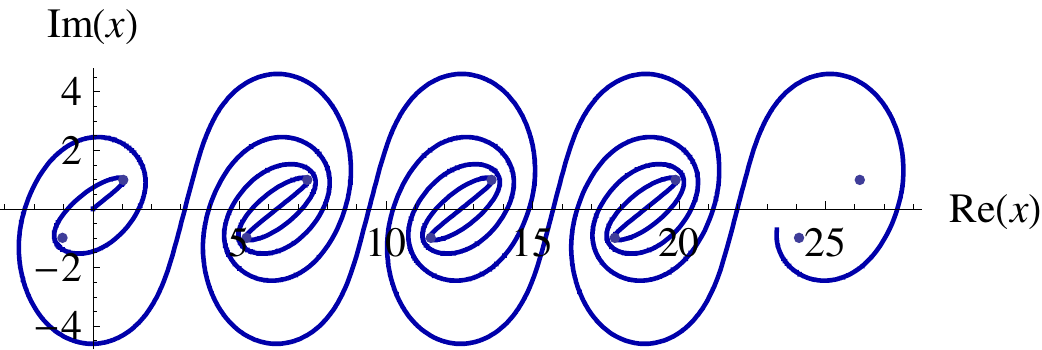}
\end{center}
\caption{Quasi-periodic trajectory of a particle with energy $E=-0.779757+i$
starting at $x=0$. For this particle the numerator $8$ in the ratio $R(E)=8/9$ 
is even, so the trajectory $x(t)$ is quasiperiodic; that is, it satisfies $x(t+T
)=2\pi+x(t)$. The dots represent the turning points, which satisfy $V(x)=E$.}
\label{F4}
\end{figure}

\begin{figure}
\begin{center}
\includegraphics[scale=1.25]{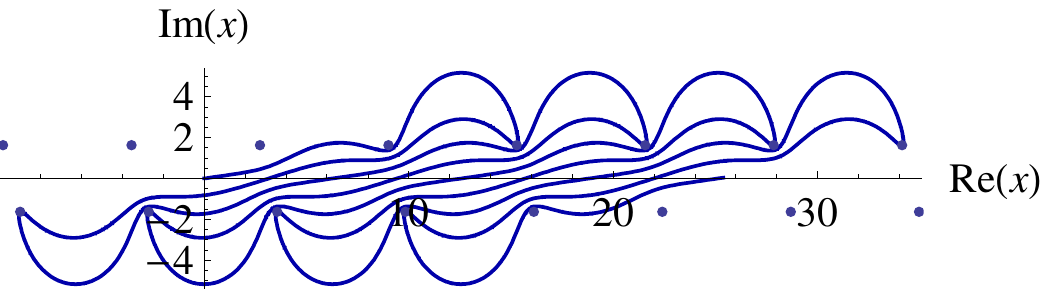}
\end{center}
\caption{Quasi-periodic trajectory $x(t)$ of a particle of energy $E=2.42891+i$
starting at $x=0$. As in Fig.~\ref{F4}, the even numerator in the ratio $R(E)=2
/9$ indicates that the trajectory is quasiperiodic.}
\label{F5}
\end{figure}

To gain a heuristic understanding of the behavior of trajectories with energies
satisfying (\ref{e9}), we use a numerical approach. We find that when $s$ is
odd, the trajectories are periodic (see Figs.~\ref{F2} and \ref{F3}). When $s$
is even, the trajectories satisfy $x(t+T)=x(t)\pm 2\pi$ (see Figs.~\ref{F4} and
\ref{F5}). Thus, our numerical work shows that we must take great care in
interpreting the inversion of the integral in (\ref{e6}).

The classical eigenenergies form a special subset of complex energies (of all
possible complex energies) and the behavior of the trajectories of particles
having these energies is independent of the initial position of the particle at
$t=0$. Had we not determined the motion of the particle analytically, it would
have been quite difficult to find energies such that the trajectory obeys $x(t+
T)-x(t)\in\{0,\pm2\pi\}$.

Let us first suppose that $E$ is real. In this case the trajectory is periodic
and it is confined to one well if $E<1$. Even if $E<-1$, where the energy of the
particle lies below the bottom of the potential well, the particle loops around
a pair of turning points (see \cite{r6}). If $E>1$, then the particle moves
with an average velocity along the positive- or negative-real $x$ axis. If $E=
1$, the particle approaches a turning point as $t\to\infty$.

\begin{figure}
\begin{center}
\includegraphics[scale=1]{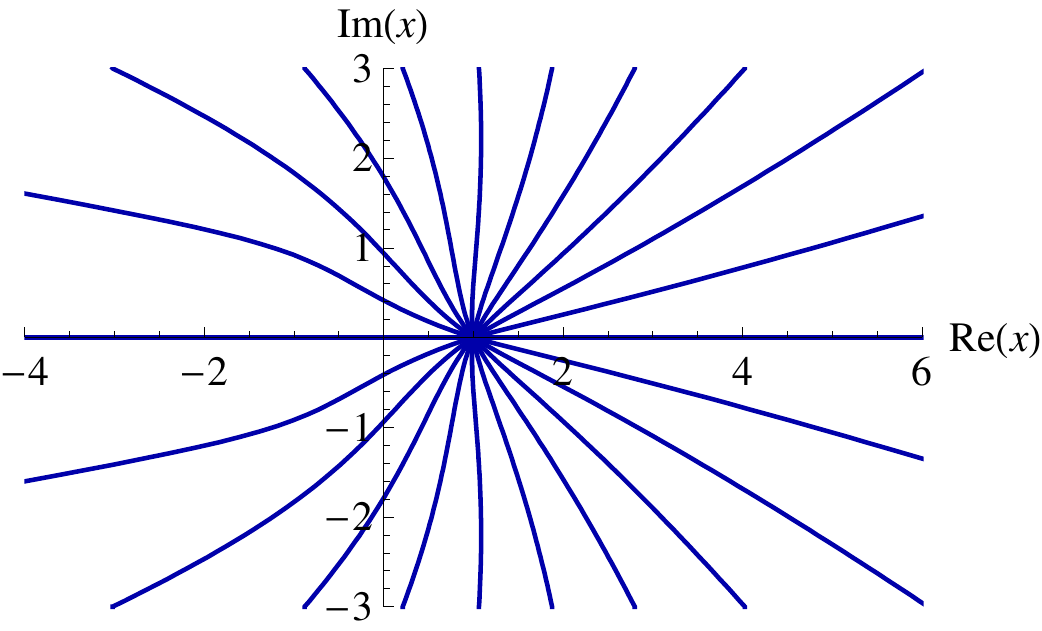}
\end{center}
\caption{A plot of the complex $E$ plane. From right to left, the curves
correspond to energies with $R(E)=1/10,\,2/10,\,\ldots,\,9/10,\,1$ in the
upper-half complex plane. In the lower half complex plane, $R(E)$ has the
opposite sign. Numerical work shows that the ratio $R(E)$ falls between $-1$ and
$1$. Because the potential is real, if $x(t)$ is a solution with energy $E$, the
complex conjugate of $x(t)$ with complex-conjugate energy $E^*$ is also a
solution. Thus, there is a symmetrical plot of eigenenergies in the lower-half
complex plane.}
\label{F6}
\end{figure}

If we solve (\ref{e9}) for complex $E$ for different rational numbers $s/r$, we
can plot the energies in the complex plane (see Fig.~\ref{F6}). Since $R(E)$ is
a continuous function almost everywhere, the relationship between the set of
special energies and the set of all complex energies is the same as the
relationship between the set of rational numbers and the set of all real
numbers; that is, the set of all special energies is dense in the complex plane,
but it is also of measure zero in the complex plane. Observe that there is a
singular point in the complex plane at $E=1$. It is not surprising that $E=1$
corresponds to a singular point because this corresponds to a particle being at
the top of the potential wells.

\section{Determining the Periodic Motion from the Ratio $s/r$}
\label{s4}

In Ref.~\cite{r7} the ratio $s/r$ was used to predict the shape of a complex
trajectory. Thus, it is natural to ask the same question here: Using just the
ratio $s/r$, what can we say about the behavior of the trajectories? For
example, the particle might move right two wells, left one well, right two
wells, left one well, and so on. We will show that the pattern of wells the
particle visits is an easily calculated function of the energy.

\begin{figure}
\begin{center}
\includegraphics[scale=1]{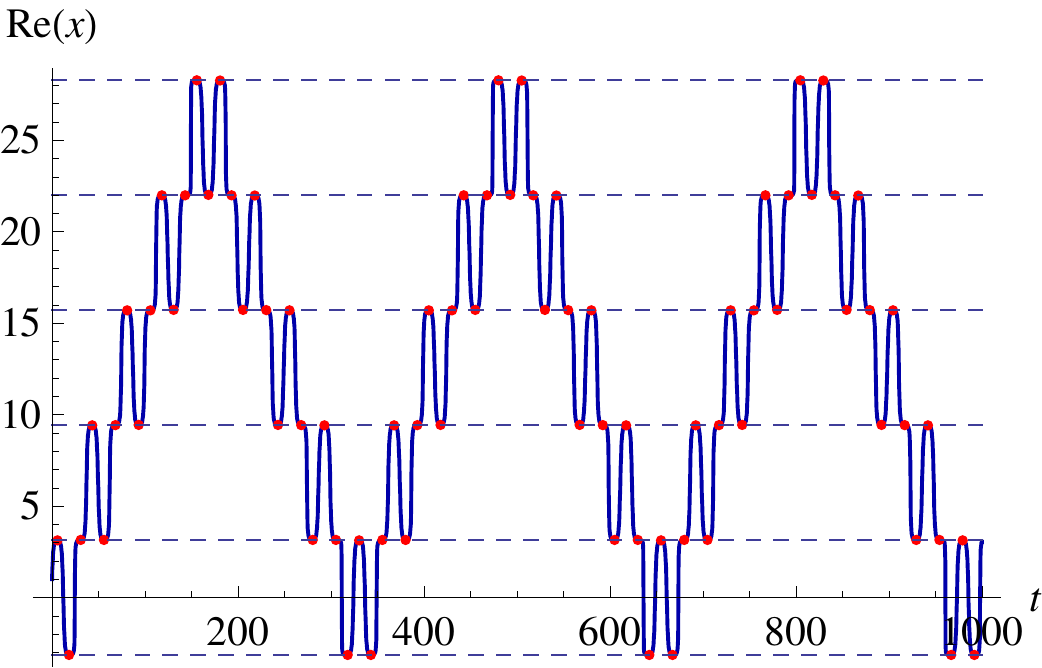}
\end{center}
\caption{Plot of ${\rm Re}\,x$ as a function of time for $E=0.999931+0.0001i$
for the ratio $9/13$ and the initial position $x(0)=1+i$. The dots (red in the
electronic version) indicate when the velocity of the particle is zero in the
${\rm Re}\,x$ direction. The dashed lines correspond to ${\rm Re}\,x=-\pi,\,\pi,
\,3\pi$, or the $x$ values that correspond to the tops of the potential $V(x)=-
\cos(x)$. The region bounded by the lowest pair of dashed lines corresponds to
the particle being in the well centered at $x=0$ with boundaries at $x=\pm\pi$.
There is a definite pattern; at each red point the particle decides whether to
reverse direction and stay in the well in which it started or to move to an
adjacent well.}
\label{F7}
\end{figure}

We consider a specific example in order to demonstrate how to determine the
pattern of motion from the ratio. In Fig.~\ref{F7} we record the motion of a
particle with ratio $9/13$ and $E=0.999931+0.0001i$ as follows: We write $+$ if
the particle continues moving in the same direction at a red dot and write $-$
if the particle changes directions at a red dot. (We begin by assuming that
${\rm Im}\,E$ is small because the pattern of hopping from well to well is
easiest to understand.) Then the pattern is
\begin{equation}
--+--+--+---+--+--+--+---+--+--+--+---
\label{e10}
\end{equation}

Note that there is a pair of minuses, a plus, a pair of minuses, a $+$, a pair
of minuses, a $+$, three minuses, a plus, and then the pattern repeats. Our goal
is to predict this pattern of plus and minus signs using just the ratio $9/13$.

We define the function 
\begin{equation}
f_\alpha(n)=\lfloor (n+1)\alpha\rfloor-\lfloor n\alpha\rfloor,
\label{e11}
\end{equation}
where $\alpha=9/13$. Then consider the sequence $\{f_\alpha(n)\}_{n=1}^\infty$.
The first few terms are as follows:
\begin{equation}
1,1,0,1,1,0,1,1,0,1,1,1,0,1,1,0,1,1,0,1,1,0,1,1,1,0,1,1,0,1,1,0,1,1,0,1,1,1,0,1.
\label{e12}
\end{equation}
If we associate $f=1$ with $-$ and $f=0$ with $+$ when we comparing (\ref{e10})
and (\ref{e12}), we find that these two sequences match exactly. Numerical work
shows that this method of determining the order in which a particle with a
particular energy visits the wells works for all energies (see Appendix).

\section{Long-Time Behavior}
\label{s5}

Having established that we can use the ratio $s/r$ to predict the pattern of
particles hopping from well to well, we can then predict the long-time behavior
of the classical particle. Whenever a particle closely approaches a boundary
between wells, it can either conduct into the next well or it can turn around
and stay in the same well. Therefore, we introduce the following measure: Let
$g_n(E)$ be the net number of wells to which the particle has moved after $n$
close approaches to boundaries between wells divided by n for a particular
energy $E$. For example, consider a case in which a particle conducts twice,
turns around twice, and the pattern repeats. If the particle is initially in the
well $[-\pi,\pi]$ and moving in the $+Re(x)$ direction, the state of the
particle, from start to finish, would progress as $([-\pi,\pi],+),\,([\pi,3\pi],
+),\,([3\pi,5\pi],+),\,([3\pi,5\pi],-),\,([3\pi,5\pi],+)$, and so on. This will
result in a net displacement of 2 wells in a total of 4 close approaches to the
boundaries between wells. So $g_n(E)$ would be $2/4=1/2$ in this example. 

If we examine $g_n(E)$ for progressively longer times (or equivalently, for more
and more times where the particle remains in the same well or conducts into an
adjacent well), the graph of $g_n(E)$ becomes more complicated. This procedure
produces a fractal-like graph where $g_\infty(E)=0$ except when the energy $E$
is such that $R(E)$ is a rational number with an even numerator.

To produce the following graphs we use $E$ to calculate the ratio $R(E)$ from
(\ref{e9}). The details of extending from rational to irrational ratios are
discussed in the Appendix. Then from $f$ we determine whether or not the
particle is going to change direction. Finally, we use this information to
calculate $g$, our measure of the motion of the particle.

\subsection{The Measure $g_n(E)$ Compared with Previous Research}
\label{ss5a}

\begin{figure}
\begin{center}
\includegraphics[scale=0.5]{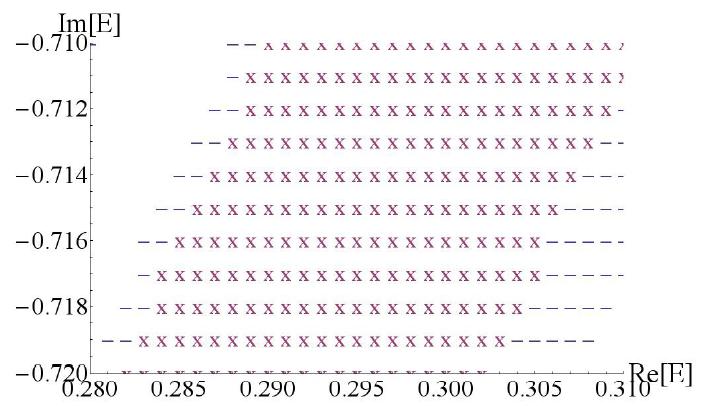}
\end{center}
\caption{This plot shows the behavior of a particle under the influence of the
periodic potential as a function of a complex energy. The X's refer to the
particle conducting, or being delocalized. The -'s refer to the particle staying
localized. On the basis of this figure one might conclude that there are sharply
defined boundaries between the energies that give rise to conducting behavior
and the energies that give rise to localized behavior. However, as explained in
this paper, this conclusion is not correct; the apparent sharpness of the band
is an artifact of observing the particle for a fixed amount of time.}
\label{F8}
\end{figure}

Using $R(E)$ to study the motion of the particles involves detailed numerical
observations, so it is helpful to show that the method described in
Sec.~\ref{s4} is consistent with previous research on the potential $V(x)=-\cos
x$. Consider Fig.~\ref{F8} (taken from Ref.~\cite{r6}). This plot indicates that
there are bands of energies that give rise to delocalized behavior.

\begin{figure}
\begin{center}
\includegraphics[scale=0.75]{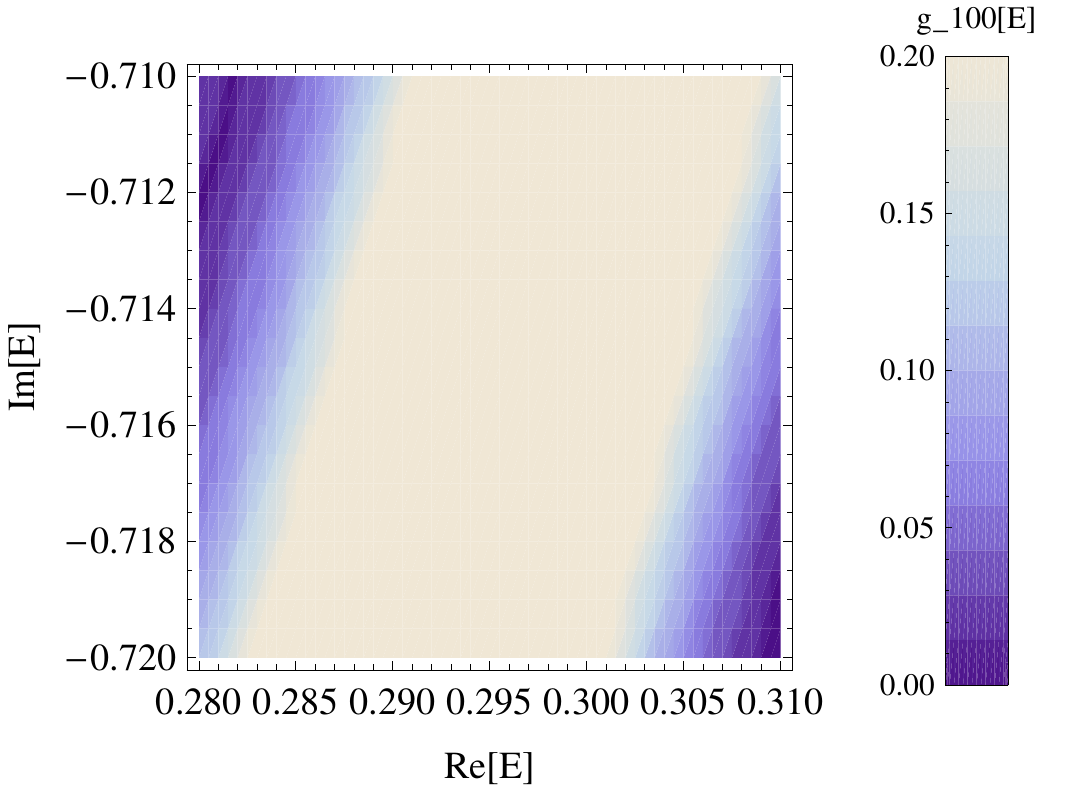}
\end{center}
\caption{Average displacement of the complex classical particle after 100 close
approaches to the boundaries between wells. The horizontal and vertical axes
give the real and imaginary parts of the energy. The intensity of each region
indicates the value of $g_{100}(E)$. The energy range is the same as in
Fig.~\ref{F8}. The plots are in good agreement; the boundaries between the
so-called conducting and localized regions match very closely.}
\label{F9}
\end{figure}

If our method is to agree  with the research reported in \cite{r6}, then the
energies that have X's in Fig.~\ref{F8} will have a nonzero value of $g_n(E)$.
For energies that have minuses in Fig.~\ref{F8}, $g_n(E)$ should be near zero.
Indeed, Fig.~\ref{F9} verifies the correspondence between the two methods. We
get an almost identical band of energies such that $g_{100}(E)=0.2$ where there
were plus signs in Fig.~\ref{F8} \cite{r11}. In the regions of minuses of
Fig.~\ref{F8}, $g_{100}(E)$ is near zero.

\begin{figure}
\begin{center}
\includegraphics[scale=0.75]{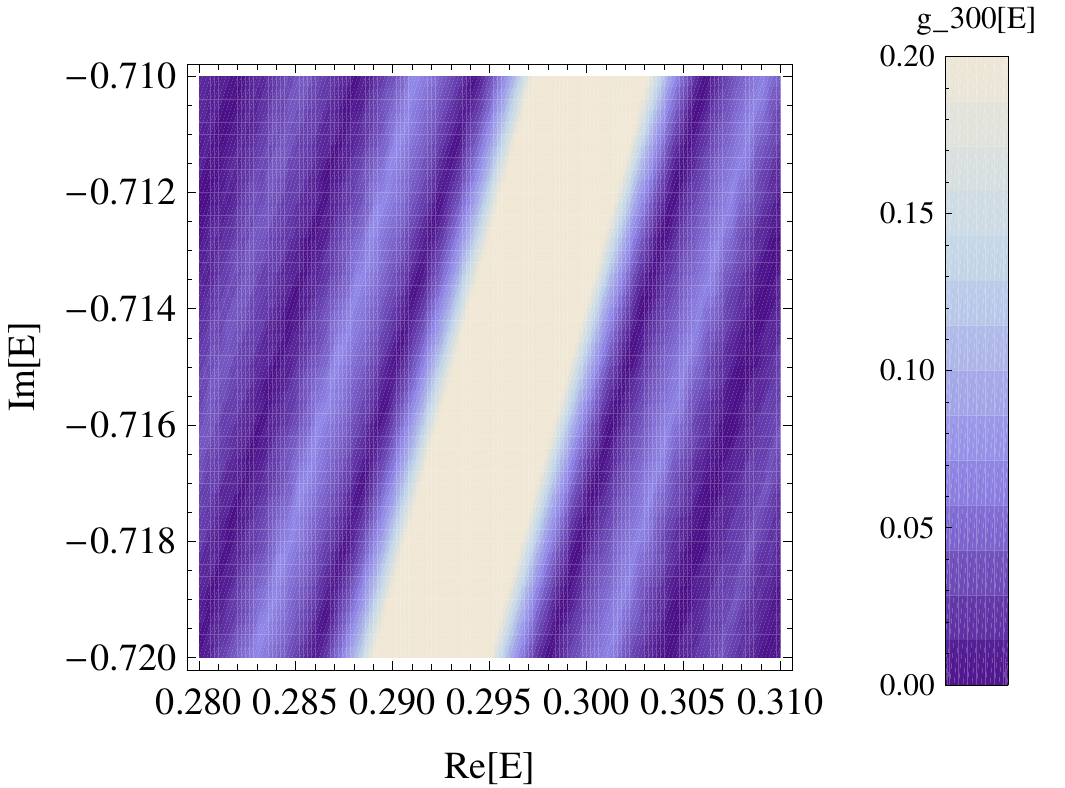}
\end{center}
\caption{Same as Fig.~\ref{F9} except that $n=300$ instead of $100$. Observe
that the center of the band still corresponds to an average displacement of one
well per five close approaches to the turning points between wells, but that the
apparent band in Fig.~\ref{F9} has fractured. If we look still further into the
future of the particle, we see that the energy band edge is not sharp.}
\label{F10}
\end{figure}

However, as discussed earlier, the dynamics are not as simple as Fig.~\ref{F8}
suggests. When we examine the long-time behavior (that is, when we measure the
average displacement for an even longer time), we find that the band fractures
into multiple bands as shown in Fig.~\ref{F10}. Yet, even though the band
divides into multiple bands, the center of both bands still corresponds to an
average displacement of one well for every five close approaches to well
boundaries.

\begin{figure}
\begin{center}
\includegraphics[scale=1]{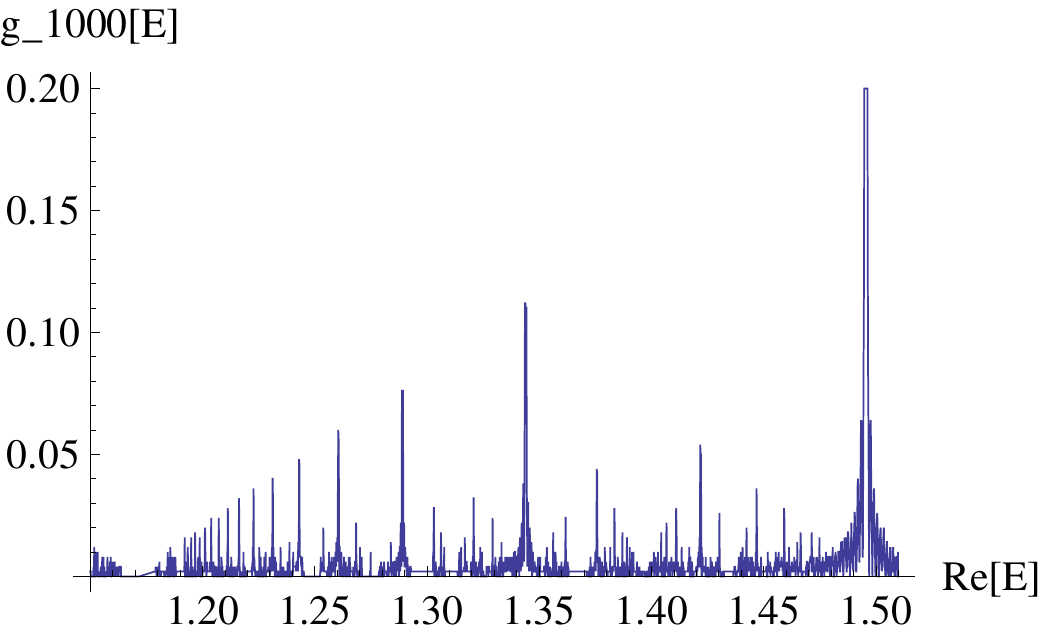}
\end{center}
\caption{A plot of $g_{1000}({\rm Re}\,E+i)$ for ${\rm Re}\,E$ ranging from
$1.15$ to $1.51$. Notice that there are no traces of bands with finite widths.
The cascade of spikes at ${\rm Re}\,E=1.49555,\,1.34399,\,1.28908,\,\ldots$
corresponds to $R(E)=2/5,\,4/9,\,6/13,\,\ldots$, where the heights of the spikes
are $1/5,\,1/9,\,1/13$, respectively. These ratios have the form $2m/(4m+1)$,
where $m$ is an integer and the height of such a spike is $1/(4m+1)$. There is
an accumulation point at ${\rm Re}\,E=1.16990$ where $R(E)=1/2$. Because this
fraction has an odd numerator, this motion is localized. Thus, we see a
transition from nonlocalized behavior (associated with an even numerator) to
localized behavior (associated with an odd numerator). This image contains
quasi-self-similar features: There is a cascade of spikes of decreasing height
and these spikes accumulate at certain energies for sequences of energies with
ratios of the form $2m/(2am+1)$ for a fixed integer $a$ and all $m$.}
\label{F11}
\end{figure}

As we let $n$ approach infinity, the bands continue to divide and we get a
fractal-like structure. In Fig.~\ref{F11} we remove the unnecessary dimension of
the imaginary part of the energy and just focus on a plot of $g$ as a function
of ${\rm Re}\,E$ where ${\rm Im}\,E$ is fixed to be 1. This figure demonstrates
how the fractal-like nature of $g_n(E)$ develops when $n$ becomes large. 

\subsection{Energy Dependence of Long-Time Behavior}
\label{ss5b}

We have demonstrated numerically that if $R(E)=\alpha=p/q$ [in the sense of
(\ref{e9})] where $p$ and $q$ are relatively prime, then $g_\infty(E)=1/q$ if
$p$ is even and $g_\infty(E)=0$ if $p$ is odd. If $\alpha$ is irrational, then
$g_\infty=0$. We examine $g_n$ as $n$ goes to infinity for $E$ such that $R(E)=
\alpha$ is irrational. To study $g_n(E)$ we approximate $\alpha$ by a suitably
close rational number. Our numerical experience shows that if we examine only up
to a fixed number $n$ of close approaches to boundaries between wells, we can
always find such a rational number $p/q$ and corresponding energy $E'$ that
reproduces the behavior demonstrated by $\alpha$ and $E$. As we choose better
rational-number approximations for $\alpha$, we note that the denominators of
such approximations approach infinity. Since $g_n(E')$ is either $0$ or $1/q$
with $q$ becoming arbitrarily large as we choose better rational approximations
for $\alpha$, $g_n(E)\to 0$ as $n\to\infty$. Thus, the particle motion is
localized in the sense that the average velocity is zero for all but a set of
measure zero of energies.

\section{Concluding Remarks}
\label{s6}

In this paper we have tried to characterize the motion of a complex classical
particle in a periodic potential. We have identified interesting trajectories
and have described the pattern of jumping from one well to another.

The results presented here are largely consistent with earlier work in which it
was observed that in the complex domain classical and quantum mechanics exhibit
many similar features. However, we have discovered new and interesting 
behaviors. Not surprisingly, when we look at the long-time behavior of a
dynamical system, we observe fractal-like behavior. We have shown that for
irrational numbers, the so-called average velocity is zero. However, this does
not eliminate the possibility that the displacement grows as a function such as
$\log t$ that is much smaller than $t$. It may be that the motion of the
particle is chaotic for energies having an irrational ratio. That is, the
particle jumps from well to well in a pattern that does not have a well-defined
asymptotic behavior. However, because the particle does not coordinate its hops
in a particular direction, the average displacement is much less than the number
of jumps of the particle.

\acknowledgments
We thank the U.S.~Department of Energy and CMB thanks the Leverhulme Foundation
for financial support. The figures in this paper were generated using
Mathematica 6.

\appendix
\setcounter{equation}{0}
\def\theequation{A\arabic{equation}}

\section{Details Concerning the Pattern}
\label{appendix}

\subsection{Motivating the Use of $f$ to Determine the Well-Jumping Pattern}
\label{ssa1}

\begin{figure}
\begin{center}
\includegraphics[scale=1]{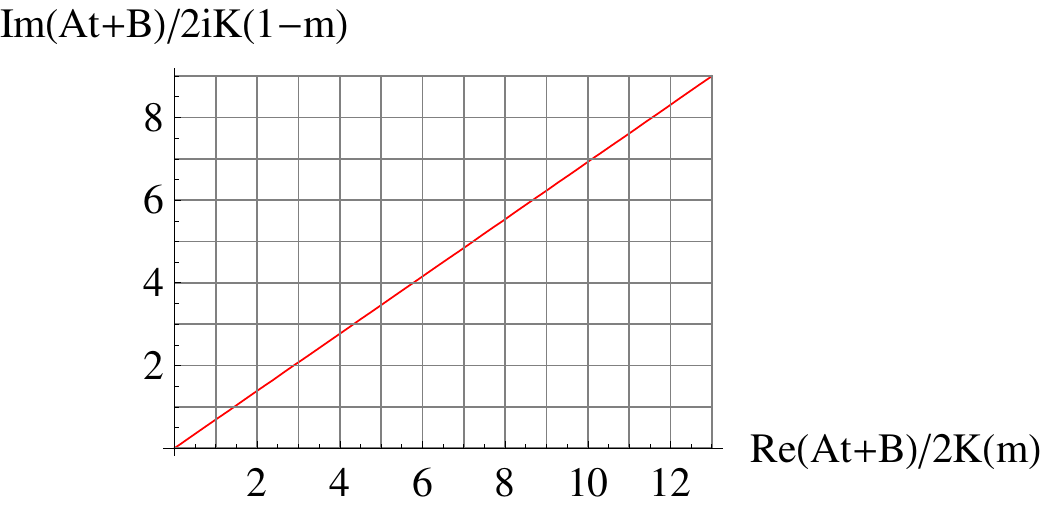}
\end{center}
\caption{The axes show the real and imaginary parts of $At+B$ scaled by the
periods in those two directions. If the energy is real, then $2K(m)$ is real and
$2iK(1-m)$ is imaginary. When $E$ becomes complex, the periods are no longer
orthogonal and the grid of rectangles shown in this figure becomes a grid of
parallelograms.}
\label{F12}
\end{figure}

Recall the form of the solution for $x(t)$ in (\ref{e7}). The function ${\rm am}
(z|m)$ is quasiperiodic in two directions in the complex-$z$ plane. These
periods are $2K(m)$ and $2iK(1-m)$. Note that $x(t)$ has the form $2{\rm am}(At+
B,m)$, where $A$ and $B$ are complex constants. So, as time evolves, $At+B$
moves through a doubly periodic domain. As a concrete example, consider a ratio
of $9/13$. The choice $R(E)=9/13$ corresponds to $At+B$ going through $13$
periods of $2K(m)$ and $9$ periods of $2iK(1-m)$ in the time interval $[0,T]$.
This is shown in Fig.~\ref{F12}.

Now we can interpret the meaning of the function $f_\alpha(n)$, as defined in
(\ref{e11}); $n\alpha$ is the $y$ value of the red line in Fig.~\ref{F12} for an
$x$ value of $n$. We can see that $\lfloor (n+1)\alpha\rfloor-\lfloor n\alpha
\rfloor$ is equal to one if the red line crosses a horizontal line between the
vertical lines $x=n$ and $x=n+1$, and is zero otherwise. That is, if we cross a
horizontal line, we have $f=1$, which corresponds to a $-$ [as in
Eq.~(\ref{e10})], and a change of direction at a red dot (as in Fig.~\ref{F7}).
So, crossing a horizontal line in Fig.~\ref{F12} should correspond to the
particle moving in the opposite direction. 

We can verify this analytically. If we move up one period [that is, $z\to z+2iK
(1-m)$], we expect the particle to be moving in the opposite direction. We can
use the following elliptic-function identities to show that there is indeed a
change in sign of $dx/dt$:
\begin{equation}
\frac{dx}{dt}=2\frac{d}{dt}{\rm am}(At+B|m)=2dn(At+B|m)(A),
\label{a1}
\end{equation}
\begin{equation}
dn(z+2iK(1-m)|m)=-dn(z|m),
\label{a2}
\end{equation}
where $dn(z|m)$ is a standard elliptic function. Although the details of taking
the functional inverse to find the form for $x(t)$ are complicated, this
calculation helps to explain why the function $f_\alpha(n)$ is relevant to this
problem.

\subsection{Numerical Justification of the use of the Pattern}

{\bf Extending from small ${\rm Im}\,E$ to any ${\rm Im}\,E$:} To identify the
hopping pattern we have assumed that the imaginary part of the energy is small
so that the boundaries of the wells are relatively well defined (as in
Fig.~\ref{F7}). However, extensive numerical work indicates that even if we let
the imaginary part of the energy be large, we still obtain the same behavior.

\begin{figure}
\begin{center}
\includegraphics[scale=0.75]{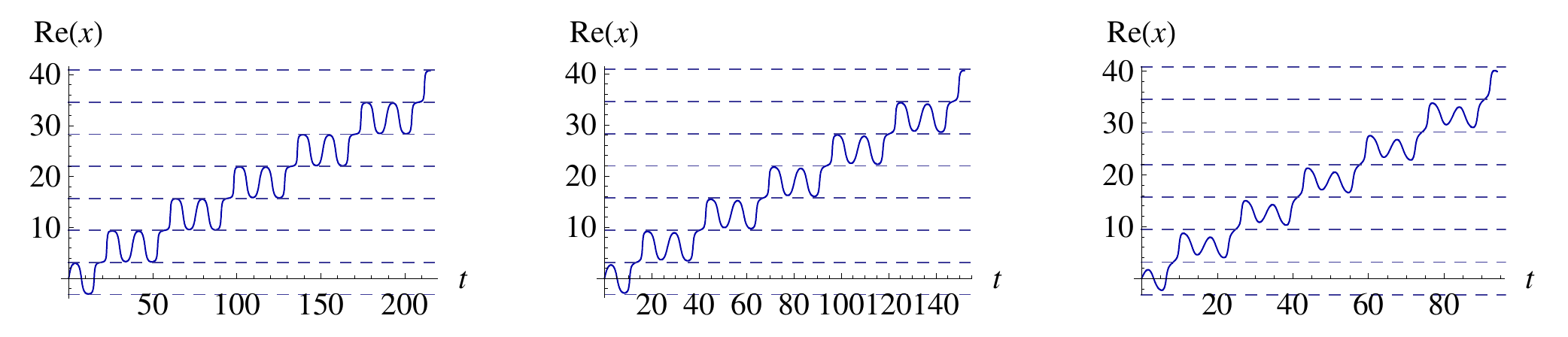}
\end{center}
\caption{Plots of ${\rm Re}\,x$ as a function of time for $x(0)=0.1i$, $R(E)=4/
5$, and varying ${\rm Im}\,E$. The dotted lines correspond to the maxima of $V(x
)$ on the real axis. From left to right, $E=0.986436+0.01i$, $0.874181+ 0.1i$,
$0.0627904+i$ and the time is taken out to $t=215,152,94$. Notice that as ${\rm
Im}\,E$ increases, the trajectories deform and the time axis deforms. These
plots demonstrate qualitatively that as ${\rm Im}\,E$ decreases, the time
required to ``tunnel'' from one well to another increases. Nevertheless, we can
see that even though the trajectories deform, the general behavior of moving
from one well to another remains the same even when ${\rm Im}\,E$ is not small.
Furthermore, these plots demonstrate that the function $g_n(E)$ is a relatively
good approximation to the average velocity as a function of energy.}
\label{F13}
\end{figure}

As the imaginary part of the energy increases, two things happen. First, the
locations of the wells become less defined. However, the pattern of moving from 
well to well remains the same, where we define the wells by the intervals $[-\pi
,\pi]$, $[\pi,3\pi]$, etc. This can be verified, for example, by plotting ${\rm
Re}\,x$ as a function of $t$ for energies that have a ratio of $9/13$ with ${\rm
Im}\,E=0.01,\,0.1,\,1$ (see Fig.~\ref{F13}).

Second, the time between close approaches to the boundaries of the wells
changes. However, this change happens slowly. Using the trajectories in
Fig.~\ref{F13}, one can also see that the time needed to move from one well to
another varies, but even for energies having small imaginary parts, this time
does not approach $\infty$ like $1/{\rm Im}\,E$. For instance, for a ratio of
$9/13$, and ${\rm Im}\,E=0.1$, this time is about $5$ and for ${\rm Im}\,E=10^{
-5}$, this time is about 10. This is contrary to previous research that had
suggested that the tunneling time is inversely proportional to the imaginary
part of the energy.

{\bf Extending from $E$ with a rational ratio to $E$ with an irrational ratio:}
We have assumed that the energy is such that $R(E)$ is rational. However,
whether or not $R(E)$ is rational is unimportant if we are tracking the particle
for a fixed amount of time because the difference between the motion for
energies that are close does not show up until a large time $t$. Thus, if we
examine the motion of a particle where $R(E)$ is irrational up to some fixed
time $t$, we can choose an energy $E'$ such that $R(E')$ is rational and such
that the two trajectories are practically identical in the time interval $[0,t]$
for any $t$.

\end{document}